\documentclass[a4paper,twoside,12pt]{article}
\input{obsjkt.sty}

\newcommand{\Msun}{\ensuremath{~{\rm M}_\odot}}                   % Solar mass symbol
\newcommand{\Rsun}{\ensuremath{~{\rm R}_\odot}}                   % Solar radius symbol
\newcommand{\rhosun}{\ensuremath{~\rho_\odot}}                    % Solar density symbol
\newcommand{\Teff}{\ensuremath{T_{\rm eff}}}                      % Effective temperature symbol
\newcommand{\Vsys}{\ensuremath{V_\gamma}}                         % Systemic velocity symbol
\newcommand{\degr}{\ensuremath{^\circ}}                           % Degree symbol
\renewcommand{\kms}{~km~s$^{-1}$}                                 % km/s symbol
\newcommand{\hip}{\textit{Hipparcos}}
\newcommand{\gaia}{\textit{Gaia}}
\newcommand{\targ}{V570~Per}
\newcommand{\targfull}{V570~Persei}

\newcommand{\Msunnom}{\hbox{$\mathcal{M}^{\rm N}_\odot$}}
\newcommand{\Rsunnom}{\hbox{$\mathcal{R}^{\rm N}_\odot$}}
\newcommand{\Lsunnom}{\hbox{$\mathcal{L}^{\rm N}_\odot$}}

\usepackage{rotating}

\begin{document} %%%%%%%%%%%%%%%%%%%%%%%%%%%%%%%%%%%%%%%%%%%%%%%%%%%%%%%%%%%%%%%%%%%%%%%%%%%%%%%%%%%%%%%%%%%%%%%%%%%%%%%%%%%%%%%%%%%%%%%%%%%%%%%%%%%%
%%%%%%%%%%%%%%%%%%%%%%%%%%%%%%%%%%%%%%%%%%%%%%%%%%%%%%%%%%%%%%%%%%%%%%%%%%%%%%%%%%%%%%%%%%%%%%%%%%%%%%%%%%%%%%%%%%%%%%%%%%%%%%%%%%%%%%%%%%%%%%%%%%%%%

\OBSheader{Rediscussion of eclipsing binaries: \targ}{J.\ Southworth}{2023 August}

\OBStitle{Rediscussion of eclipsing binaries. Paper XIV. \\ The F-type system V570 Persei}

\OBSauth{John Southworth}

\OBSinstone{Astrophysics Group, Keele University, Staffordshire, ST5 5BG, UK}

% \OBSabstract{Abs abs abs abs abs abs abs abs abs abs abs abs abs abs abs abs abs abs abs abs abs abs abs abs abs abs abs abs abs abs abs abs abs abs abs abs abs abs abs abs abs abs abs abs abs abs abs abs abs abs abs abs abs abs abs abs abs abs abs abs abs abs abs abs abs abs abs abs abs abs abs abs abs abs abs abs abs abs abs abs abs abs abs abs abs abs abs abs abs abs abs abs abs abs abs abs abs abs abs abs abs abs abs abs abs abs abs abs abs abs abs abs abs abs abs abs abs abs abs abs abs abs abs abs abs abs abs abs abs abs abs abs abs abs abs abs abs abs abs abs abs abs abs abs abs abs abs abs abs abs abs abs abs abs abs abs abs abs abs abs abs abs abs abs abs abs abs abs abs abs abs abs abs abs abs abs abs abs abs abs abs abs abs abs abs abs abs abs abs abs abs abs abs abs abs abs abs abs abs abs abs abs abs abs abs abs abs abs abs abs abs abs abs abs abs abs abs abs abs abs abs abs abs abs abs abs abs abs abs abs abs abs abs abs abs abs abs abs abs abs abs abs abs.}

\OBSabstract{\targ\ is a binary star system containing two F-type stars in a 1.90~d period circular orbit. It shows shallow partial eclipses that were discovered from its \hip\ light curve. We present an analysis of this system based on two sectors of high-quality photometry from the NASA Transiting Exoplanet Survey Satellite (TESS) mission, and published spectroscopic light ratio and radial velocity measurements. We find masses of $1.449 \pm 0.006$ and $1.350 \pm 0.006$\Msun, and radii of $1.538 \pm 0.035$ and $1.349 \pm 0.032$\Rsun. The radius measurements are set by the spectroscopic light ratio and could be improved by obtaining a more precise light ratio. The eclipses in the TESS data arrived $660 \pm 30$~s later than expected, suggesting the presence of a faint third body on a wider orbit around the eclipsing system. Small trends in the residuals of the fit to the TESS lightcurve are attributed to weak starspots. The distance to the system is close to the \gaia\ DR3 value, but the \gaia\ spectroscopic orbit is in moderate disagreement with the results from the published ground-based data.}

%%%%%%%%%%%%%%%%%%%%%%%%%%%%%%%%%%%%%%%%%%%%%%%%%%%%%%%%%%%%%%%%%%%%%%%%%%%%%%%%%%%%%%%%%%%%%%%%%%%%%%%%%%%%%%%%%%%%%%%%%%%%%%%%%%%%%%%%%%%%%%%%%%%%%

\section*{Introduction}

Detached eclipsing binary stars (dEBs) are our main source of measurements of the physical properties of normal stars. The number of dEBs for which precise measurements are available is increasing gradually, as traced by reviews of this subject \cite{Popper80araa,Andersen91aarv,Torres++10aarv} as well as compiled catalogues \cite{Malkov+06aa,Eker+14pasa,Me15aspc}. The Detached Eclipsing Binary Catalogue (DEBCat\footnote{\texttt{https://www.astro.keele.ac.uk/jkt/debcat/}}, ref.~\cite{Me15aspc}) currently lists just over 300 dEBs for which masses and radii are measured to 2\% precision or better, helped by the widespread availability of light curves from space telescopes \cite{Me21univ}.

dEBs are useful in understanding the physical processes that govern the structure and evolution of stars. They have been used to calibrate the amount of convective core overshooting \cite{Andersen++90apj,ClaretTorres16aa,ClaretTorres18apj} albeit with conflicting results \cite{ConstantinoBaraffe18aa}, the size of the convective core in massive stars \cite{Tkachenko+20aa}, mixing length \cite{Graczyk+16aa}, and the radii of low-mass stars \cite{Torres13an,Jennings+23mn}. They are also sources of distance measurements which have been used to calibrate the cosmological distance scale \cite{Pietrzynski+19nat,Freedman+20apj}

We are currently pursuing a project to increase the number of dEBs with reliable measurements of their masses and radii \cite{Me20obs}, primarily using new observations from the NASA Transiting Exoplanet Survey Satellite (TESS) mission \cite{Ricker+15jatis}. TESS has observed thousands of dEBs \cite{IJspeert+21aa,JustesenAlbrecht21apj,Prsa+21apjs}, many of which have available high-quality radial velocity (RV) measurements. In this context, we present an analysis of the \targfull\ system.

\targ\ (Table\,\ref{tab:info}) is an F-type dEB which was discovered using data from the \hip\ satellite \cite{Vanleeuwen+97aa} and given its variable-star name by Kazarovets et al.\ \cite{Kazarovets+99ibvs}. It was selected for analysis by Munari et al.\ \cite{Munari+01aa} in the context of assessing the expected performance of the \gaia\ satellite in the study of dEBs. These authors used the \hip\ photometry of \targ\ along with ground-based spectroscopy restricted to the 850--875~nm wavelength range to mimic the expected characteristics of the \gaia\ observations. They measured the masses of the components of \targ\ to 2.5\%, and the radii to low precisions of 10\% and 25\% due to the large scatter in the \hip\ data and the shallow eclipses shown by this dEB. Tomasella et al.\ \cite{Tomasella+08aa2} (hereafter T08) presented a more detailed study of \targ\ based on new ground-based photometry, and the same spectroscopy but this time using the full available 450--948~nm wavelength range. They constrained the model of the light curve using spectroscopically-measured light contributions of the two stars in the $V$-band. They determined the atmospheric parameters of the component stars via a $\chi^2$ fit of synthetic spectra to their observed spectra, a process which neglected the systematic errors inherent in this method.

% ok /home/jkt/papers/writtenup/2008A+A...483..263T.pdf

\begin{table}[t]
\caption{\em Basic information on \targ. \label{tab:info}}
\centering
\begin{tabular}{lll}
{\em Property}                            & {\em Value}                 & {\em Reference}                   \\[3pt]
Right ascension (J2000)                   & 03:09:34.94                 & \cite{Gaia21aa}                   \\
Declination (J2000)                       & +48:38:28.7                 & \cite{Gaia21aa}                   \\
% Bright Star Catalogue                   & HR NNNN                     & \cite{HoffleitJaschek91}          \\
Henry Draper designation                  & HD 19457                    & \cite{CannonPickering18anhar}     \\
\textit{Hipparcos} designation            & HIP 1673                    & \cite{Hipparcos97}                \\
% \textit{Tycho} designation              & TYC NNNN-NNN-N              & \cite{Hog+00aa}                   \\
\textit{Gaia} DR3 designation             & 435997252803241856          & \cite{Gaia21aa}                   \\
\textit{Gaia} DR3 parallax                & $8.2952 \pm 0.0355$ mas     & \cite{Gaia21aa}                   \\          % d = 120.55 +/- 0.52 pc
TESS\ Input Catalog designation           & TIC 116991977               & \cite{Stassun+19aj}               \\
$B$ magnitude                             & $8.55 \pm 0.02$             & \cite{Hog+00aa}                   \\          % \cite{Henden+15aas} for APASS
$V$ magnitude                             & $8.09 \pm 0.01$             & \cite{Hog+00aa}                   \\          % \cite{Hog+00aa} for Tycho
$J$ magnitude                             & $7.160 \pm 0.026$           & \cite{Cutri+03book}               \\
$H$ magnitude                             & $6.948 \pm 0.017$           & \cite{Cutri+03book}               \\
$K_s$ magnitude                           & $6.882 \pm 0.020$           & \cite{Cutri+03book}               \\
Spectral type                             & F3 V + F5 V                 & \cite{Tomasella+08aa2}            \\[3pt]
\end{tabular}
\end{table}

%%%%%%%%%%%%%%%%%%%%%%%%%%%%%%%%%%%%%%%%%%%%%%%%%%%%%%%%%%%%%%%%%%%%%%%%%%%%%%%%%%%%%%%%%%%%%%%%%%%%%%%%%%%%%%%%%%%%%%%%%%%%%%%%%%%%%%%%%%%%%%%%%%%%%

\section*{Observational material}

\begin{figure}[t] \centering \includegraphics[width=\textwidth]{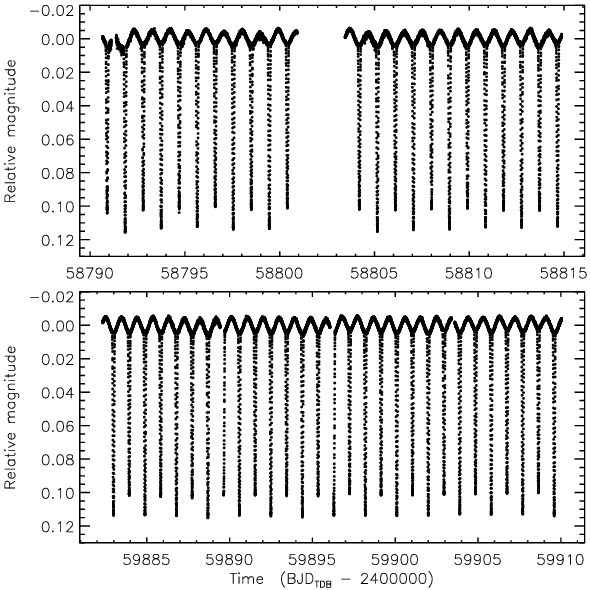} \\
\caption{\label{fig:time} TESS\ short-cadence SAP photometry of \targ\ from sectors
18 (top), and 58 (bottom). The flux measurements have been converted to magnitude
units then rectified to zero magnitude by subtraction of the median.} \end{figure}

% Sector 18 (2019-Nov-02 to 2019-Nov-27, in cycle 2): observed in camera 1.
% Sector 58 (2022-Oct-29 to 2022-Nov-26, in cycle 5): observed in camera 1.

The TESS mission \cite{Ricker+15jatis} observed \targ\ in sectors 18 (2019/11/02 to 2019/11/27) and 58 (2022/10/29 to 2022/11/26), in both cases in short cadence mode with a 120~s sampling rate. We used the {\sc lightkurve} package \cite{Lightkurve18} to download these data and reject points flagged as bad. The simple aperture photometry (SAP) and pre-search data conditioning SAP (PDCSAP) data \cite{Jenkins+16spie} are almost indistinguishable, so we used the SAP data in our analysis for consistency with previous papers in this series.

We converted the data to differential magnitude and subtracted the median magnitude for further analysis, ending up with 15\,256 datapoints from sector 18 and 19\,475 from sector 58. On further inspection we found that the first stretches of data from both halves of the sector 18 light curve were affected by instrumental systematics, so we trimmed them by removing data in the intervals [2458790.6,2458792.5] and [2458801.0,2458804.7]. This left a total of 32\,719 datapoints over both TESS sectors (Fig.~\ref{fig:time}).

We queried the \gaia\ DR3 database\footnote{\texttt{https://vizier.cds.unistra.fr/viz-bin/VizieR-3?-source=I/355/gaiadr3}} for objects within 2~arcmin of \targ. A total of 108 were found, all of which are fainter than \targ\ by at least 7.2~mag in the \gaia\ $G$ band. We deduce that the amount of light contaminating the TESS aperture for this dEB is negligible.

%%%%%%%%%%%%%%%%%%%%%%%%%%%%%%%%%%%%%%%%%%%%%%%%%%%%%%%%%%%%%%%%%%%%%%%%%%%%%%%%%%%%%%%%%%%%%%%%%%%%%%%%%%%%%%%%%%%%%%%%%%%%%%%%%%%%%%%%%%%%%%%%%%%%%

\section*{Light curve analysis}

\begin{figure}[t] \centering \includegraphics[width=\textwidth]{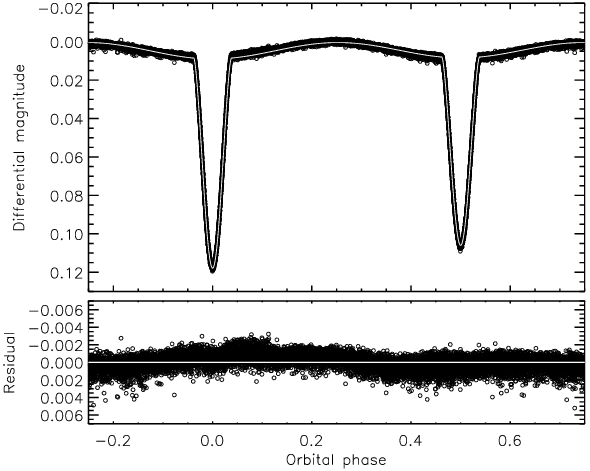} \\
\caption{\label{fig:phase18} Best fit to the TESS sector 18 light curve of \targ\
using {\sc jktebop} as a function of orbital phase. The residuals are shown on an
enlarged scale in the lower panel.} \end{figure}

\begin{figure}[t] \centering \includegraphics[width=\textwidth]{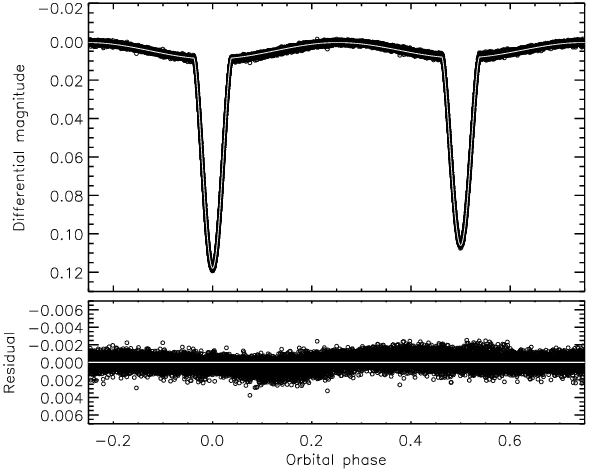} \\
\caption{\label{fig:phase58} Best fit to the TESS sector 58 light curve of \targ\
using {\sc jktebop} as a function of orbital phase. The residuals are shown on an
enlarged scale in the lower panel.} \end{figure}

We modelled the light curves from the two sectors both individually and together, using version 43 of the {\sc jktebop}\footnote{\texttt{http://www.astro.keele.ac.uk/jkt/codes/jktebop.html}} code \cite{Me++04mn2,Me13aa}. In all cases the parameters of the fit included the fractional radii of the stars ($r_{\rm A}$ and $r_{\rm B}$), expressed as their sum ($r_{\rm A}+r_{\rm B}$) and ratio ($k = {r_{\rm B}}/{r_{\rm A}}$), the orbital inclination ($i$), the central surface brightness ratio ($J$), the ephemeris (period $P$ and reference time of primary minimum $T_0$) and the coefficients of the reflection effect. We define star~A to be the one eclipsed at the deeper minimum and star~B to be its companion. A circular orbit was assumed based on the appearance of the light curve and of the RVs presented by T08 -- when allowing for an eccentric orbit we found a best-fitting eccentricity of $e=0.0053$ and almost no change in the other parameters. We included a quadratic function versus time for each half-sector to account for slow changes in the brightness of the dEB due to instrumental effects.

The eclipses are partial and shallow, so the light curve solution suffers from a strong degeneracy between $k$, $i$ and $J$ (e.g.\ refs.\ \cite{Torres+00aj} and \cite{Me++07aa}). This effect was found by T08 when modelling their ground-based photometry, and remains present in the much more extensive and higher-precision TESS data used in the current study. We therefore applied a spectroscopic light ratio as a constraint, in the same way as done in our work on V1022~Cas \cite{Me21obs2} and HD~23642 \cite{Me++23mn}. The light contributions found by T08 correspond to a light ratio of $\ell_{\rm B}/\ell_{\rm A} = 0.667 \pm 0.053$ in the $V$-band. We propagated this to the TESS passband using the response function from Ricker et al.\ \cite{Ricker+15jatis}, theoretical spectra from Allard et al.\ \cite{Allard++12rspta}, and the effective temperature (\Teff) values from T08, finding $\ell_{\rm B}/\ell_{\rm A} = 0.703 \pm 0.057$.

Limb darkening (LD) was included in the fit \cite{Me23obs2} using the power-2 law \cite{Hestroffer97aa} and theoretical LD coefficients \cite{ClaretSouthworth22aa}. Fitting for the scaling coefficient (``c'' in the terminology of Maxted \cite{Maxted18aa}) for both stars yielded determinate values and little change in the other parameters, so was adopted as the default approach.

The amount of third light ($L_3$) has a significant effect on the best-fitting parameter values. If fitted, it converges to a formally significant but unphysically negative value ($-0.083 \pm 0.018$) despite the negligible amount of light from nearby stars (see previous section). We therefore fixed it at zero in our default solution, but added contributions to the errorbars based on the change in parameter values by assuming $L_3 = 2\%$ instead. For information, such an assumption decreases $r_{\rm A}$ by 1.1\% and increases $r_{\rm B}$ by 0.4\%.

%                                                 r1                      r2
% Lrat=0.67
% Standard solution:                              0.1735 +/-  0.0024      0.1455 +/-  0.0031
% LD fitted                                       0.1718 +/-  0.0025      0.1466 +/-  0.0031
% LD fix but perturbed ("2"):                     0.1735 +/-  0.0135      0.1455 +/-  0.0181
% L3 fitted:                                      0.1787 +/-  0.0028      0.1454 +/-  0.0033      L3 = -0.083 +/- 0.018
% ecosw and esinw fitted:                         0.1760 +/-  0.0027      0.1419 +/-  0.0027      e = 0.0053 +/- 0.0006
% Lrat=0.703:
% Standard solution:                              0.1728                  0.1463
% L3 = 0.02 fixed                                 0.1710                  0.1469

The best fits to the light curves from the two sectors are shown in Figs.\ \ref{fig:phase18} and \ref{fig:phase58}. These plots show the result of a fit to both sectors simultaneously, but divided into individual sectors in the plots. Slow trends in the residuals are apparent in both cases, and are discussed below.

The fitted parameters are given in Table~\ref{tab:jktebop}. Uncertainties in the parameters were determined using Monte Carlo and residual-permutation simulations \cite{Me++04mn,Me08mn}. The Monte Carlo errorbars are significantly larger than the residual-permutation alternatives because the latter do not account for the uncertainty in the spectroscopic light ratio. We therefore adopted the Monte Carlo errorbars for all parameters. The dominant source of uncertainty is the spectroscopic light ratio, which could be improved by further observations and analysis.

\begin{table} \centering
\caption{\em \label{tab:jktebop} Adopted parameters of \targ\ measured from the
TESS\ light curves using the {\sc jktebop} code. The uncertainties are 1$\sigma$
and were determined using Monte Carlo and residual-permutation simulations.}
\begin{tabular}{lc}
{\em Parameter}                           &       {\em Value}                 \\[3pt]
{\it Fitted parameters:} \\
Time of primary eclipse (BJD$_{\rm TDB}$) & $ 2459894.392999 \pm  0.000009   $ \\
Orbital period (d)                        & $     1.90093830 \pm  0.00000002 $ \\
Orbital inclination (\degr)               & $      77.294    \pm  0.048      $ \\
Sum of the fractional radii               & $       0.31715  \pm  0.00057    $ \\
Ratio of the radii                        & $       0.877    \pm  0.036      $ \\
Central surface brightness ratio          & $       0.8767   \pm  0.0033     $ \\
% Third light                             & $       0.0001   \pm  0.0008     $ \\
LD coefficient $c$ for star~A             & $       0.548    \pm  0.017      $ \\
LD coefficient $c$ for star~B             & $       0.516    \pm  0.020      $ \\
LD coefficient $\alpha$ for star~A        &             0.498 (fixed)         \\
LD coefficient $\alpha$ for star~B        &             0.467 (fixed)         \\
Orbital eccentricity                      &             0.0~~ (fixed)         \\
{\it Derived parameters:} \\
Fractional radius of star~A               & $       0.1690   \pm  0.0028     $ \\
Fractional radius of star~B               & $       0.1482   \pm  0.0035     $ \\
Light ratio $\ell_{\rm B}/\ell_{\rm A}$   & $       0.683    \pm  0.060      $ \\[3pt]
\end{tabular}
\end{table}

%%%%%%%%%%%%%%%%%%%%%%%%%%%%%%%%%%%%%%%%%%%%%%%%%%%%%%%%%%%%%%%%%%%%%%%%%%%%%%%%%%%%%%%%%%%%%%%%%%%%%%%%%%%%%%%%%%%%%%%%%%%%%%%%%%%%%%%%%%%%%%%%%%%%%

\section*{The out-of-eclipse variability}

The best fits to the light curves (Figs.\ \ref{fig:phase18} and \ref{fig:phase58}) show slow trends in the residuals which differ between the two sectors. Our preferred interpretation of this is small brightness variations present on the surface of one or both stars, with the star(s) rotating synchronously with the orbit in order to obtain the consistent phasing in Figs.\ \ref{fig:phase18} and \ref{fig:phase58}. This could be caused by starspots, and evolution of the spot configuration is a natural explanation for the differences between the residuals of the fits to the two sectors. The \Teff\ values of the stars are relatively high for this explanation, but are only slightly higher than KIC 5359678 for which spot activity was clearly detected \cite{Wang+21mn,Me22obs3}. The lack of increased residuals during eclipse suggests the spots are either a similar temperature to the rest of the photosphere and/or are located on parts of the star(s) that are not eclipsed.

We checked for the possibility of pulsations by calculating a periodogram of the residuals of the fit to the data from sector 58, using the {\sc period04} code \cite{LenzBreger04iaus}. Significant signals were found at the orbital period and half the orbital period, in agreement with the starspot hypothesis. No evidence for either $\delta$~Scuti or $\gamma$~Doradus pulsations were found, despite a significant number of such pulsators now being known in dEBs \cite{GaulmeGuzik19aa,Me21obs6,MeVanreeth22mn,Kahraman+22raa,Chen+22apjs}.

%%%%%%%%%%%%%%%%%%%%%%%%%%%%%%%%%%%%%%%%%%%%%%%%%%%%%%%%%%%%%%%%%%%%%%%%%%%%%%%%%%%%%%%%%%%%%%%%%%%%%%%%%%%%%%%%%%%%%%%%%%%%%%%%%%%%%%%%%%%%%%%%%%%%%

\section*{Radial velocities}

T08 measured RVs of both stars from each of 31 high-quality \'echelle spectra obtained using the Asiago 1.8~m telescope. We obtained these from table~2 in T08 and modelled them using {\sc jktebop}, adopting a circular orbit and separate systemic velocities (\Vsys) for the two stars. We fitted for velocity amplitudes ($K_{\rm A}$ and $K_{\rm B}$), ${\Vsys}_{\rm ,A}$, ${\Vsys}_{\rm ,B}$ and $T_0$. The period was fixed at the value from Table~\ref{tab:jktebop}. Uncertainties were calculated from 1000 Monte Carlo simulations \cite{Me++04mn2,Me21obs5} after adjusting the sizes of the errorbars to give a reduced $\chi^2$ of unity for the RVs for each star.

\begin{figure}[t] \centering \includegraphics[width=\textwidth]{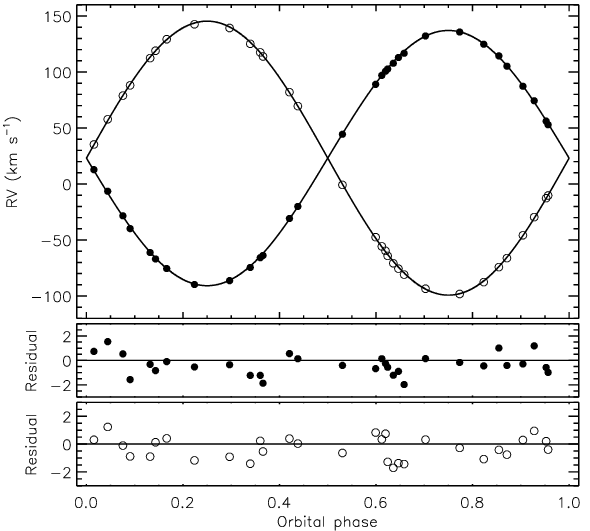} \\
\caption{\label{fig:rv} RVs of \targ\ from T08 (filled circles for star~A and
open circles for star~B) compared to the best-fitting spectroscopic orbits
from our own analysis using {\sc jktebop} (solid curves). The residuals are
given in the lower panels separately for the two components.} \end{figure}

We found $K_{\rm A} = 113.94 \pm 0.24$\kms, $K_{\rm B} = 122.33 \pm 0.22$\kms, ${\Vsys}_{\rm ,A} = 23.15 \pm 0.16$\kms\ and ${\Vsys}_{\rm ,B} = 23.09 \pm 0.14$\kms, where the uncertainties in the systemic velocities do not include any transformation onto a standard system. The best fits are shown in Fig.~\ref{fig:rv}. We cannot compare the $K_{\rm A}$ and $K_{\rm B}$ values directly with the results from T08 because they did not calculate these parameters explicitly.

% TESS T0: 59894.392999 (9)
% RV mean time: 51466.4
% RV eclipse time close to mean time: 51465.633
% Cycle difference: 4434
% Uncertainty from P: 0.0011084999
% Uncertainty from T0: 0.000009
% Uncertainty total: 0.0011085365
% Diff in timings: 59894.392999d0-59894.385378d0 = 0.00762 d = 658 s
% T0 uncertainty from RV fit: 0.00034
% T0 uncertainty from ephemeris: 0.00111
% T0 uncertainty total: 0.00117
% T0 RV errorbar: 0.00034
% Offset significance: 3.4 sigma

We found an offset of $658 \pm 29$~s between the $T_0$ from the RV fit and that predicted from the ephemeris in Table~\ref{tab:jktebop}. Further investigation suggests that this offset is also present in the times of minimum light given by T08 and Hubscher et al.\ \cite{Hubscher+10ibvs}. As the current work is the first by the author that used the {\sc lightkurve} package to access TESS data, one possibility is that this approach has caused an offset in the timestamps. We checked this by using {\sc lightkurve} to download TESS light curves for ZZ~UMa and ZZ~Boo and compared them to those used in refs.\ \cite{Me23obs1} and \cite{Me22obs6}. No offset in the timings was found, suggesting that the timing offset is an astrophysical effect, perhaps caused by a third component on a wider orbit around \targ.

\targ\ is present in the \gaia\ DR3 \textit{Non-single-star orbital models for sources compatible with Double Lined Spectroscopic binary model} catalogue\footnote{\texttt{https://vizier.cds.unistra.fr/viz-bin/VizieR-3?-source=I/357/tbosb2}} which reports objects detected as double-lined and with a fitted spectroscopic orbit \cite{Gaia16aa,Gaia22aa}. The orbital parameters given are $e = 0.0029 \pm 0.0019$, $K_1 = 123.86 \pm 0.28$\kms\ and $K_2 = 113.82 \pm 0.24$\kms, based on RVs from 24 spectra. The eccentricity is very small and consistent with zero, as expected. We find that $K_2$ is in good agreement with our $K_{\rm A}$, but that $K_1$ is moderately discrepant with our $K_{\rm B}$. It is clear that the identities of the stars have been swapped, but the source of the $K_1$/$K_{\rm B}$ discrepancy is unknown. We chose not to use these results because the spectra and RVs on which they are based are not publicly available so cannot be checked. It is relevant that Tokovinin \cite{Tokovinin23aj} has found issues with the \gaia\ DR3 $K_1$ and $K_2$ values in the sense that a significant fraction (14 of 22 in that case) have underestimated values or other problems.

\section*{Physical properties of \targ}

We determined the physical properties of \targ\ using the {\sc jktabsdim} code \cite{Me++05aa}. The input values to this were: (i) the $r_{\rm A}$, $r_{\rm B}$, $i$ and $P$ from Table~\ref{tab:jktebop}; the $K_{\rm A}$ and $K_{\rm B}$ from the RV analysis; the \Teff\ values from T08 with the errorbars increased to $\pm$50~K to account for the systematic uncertainties of the \Teff\ scale for F-stars \cite{Depascale+14aa,Ryabchikova+16mn,Jofre++19araa}; an interstellar reddening of $E(B-V) = 0.05 \pm 0.02$~mag from the {\sc stilism}\footnote{\texttt{https://stilism.obspm.fr}} online tool \cite{Lallement+14aa,Lallement+18aa}; the $B$ and $V$ magnitudes from Tycho-2 \cite{Hog+00aa} which are averages of 12 measurements at effectively random orbital phases; and the $JHK_s$ magnitudes from 2MASS \cite{Cutri+03book} converted to the Johnson system using the transformations from Carpenter \cite{Carpenter01aj}. The 2MASS magnitudes were taken at phase 0.10 so are representative of the average brightness of the system. The results are given in Table~\ref{tab:jktebop}, where the errorbars have been propagated individually from each input parameter.

\begin{table} \centering
\caption{\em Physical properties of \targ\ defined using the nominal solar units given by IAU
2015 Resolution B3 (ref.\ \cite{Prsa+16aj}). \label{tab:absdim}}
\begin{tabular}{lr@{\,$\pm$\,}lr@{\,$\pm$\,}l}
{\em Parameter}        & \multicolumn{2}{c}{\em Star A} & \multicolumn{2}{c}{\em Star B}    \\[3pt]
Mass ratio   $M_{\rm B}/M_{\rm A}$          & \multicolumn{4}{c}{$0.9314 \pm 0.0026$}       \\
Semimajor axis of relative orbit (\Rsunnom) & \multicolumn{4}{c}{$9.100 \pm 0.013$}         \\
Mass (\Msunnom)                             &  1.4489 & 0.0063      &  1.3495 & 0.0062      \\
Radius (\Rsunnom)                           &  1.538  & 0.035       &  1.349  & 0.032       \\
Surface gravity ($\log$[cgs])               &  4.225  & 0.020       &  4.308  & 0.021       \\
Density ($\!\!$\rhosun)                     &  0.398  & 0.027       &  0.550  & 0.039       \\
Synchronous rotational velocity ($\!\!$\kms)& 40.93   & 0.92        & 35.89   & 0.85        \\
Effective temperature (K)                   &   6842  & 50          &   6562  & 50          \\
Luminosity $\log(L/\Lsunnom)$               &   0.669 & 0.023       &   0.483 & 0.024       \\
$M_{\rm bol}$ (mag)                         &   3.068 & 0.058       &   3.533 & 0.061       \\
Distance (pc)                               & \multicolumn{4}{c}{$117.2 \pm 2.3$}           \\[3pt]
\end{tabular}
\end{table}

% IDL> print, [0.035,0.032,0.0063,0.0062]/[1.538,1.349,1.4489,1.3495]*100
%       2.27568      2.37213     0.434813     0.459429

The agreement between the measurements in Table~\ref{tab:jktebop} and the results from T08 is good, with all quantities within 1$\sigma$. The radii of the stars have been determined to 2.3\% precision, which is slightly worse than managed by T08 despite the availability of much better photometry for the current study. This arises because the precision of the radius measurements is limited by the spectroscopic light ratio applied in the photometric analysis, and perhaps from underestimated errorbars in T08. A better spectroscopic light ratio is needed to measure the radii more precisely.

The synchronous rotational velocities are consistent with the $v \sin i$ values measured by T08. This is in agreement with our assertion that the trends in the residuals of the fit to the light curves are due to starspots rotating synchronously with the orbit.

Inversion of the \gaia\ DR3 parallax gives a distance to the system of $d = 120.55 \pm 0.52$~pc, which is 1.4$\sigma$ longer than that found in our own work via the $K$-band surface brightness method \cite{Me++05aa} and calibrations from Kervella et al.\ \cite{Kervella+04aa}. An increase in $E(B-V)$ to 0.1~mag would bring our optical ($BV$) and infrared ($JHK_s$) distances into better agreement at the expense of shortening the distance measurement to $115.8 \pm 2.3$~pc; this reddening is significantly more than the $0.023 \pm 0.007$~mag found by T08 from the interstellar sodium and potassium lines. The shorter distance could then be compensated by adopting larger \Teff\ values for the stars. The \gaia\ distance is questionable because the renormalised unit weight error (RUWE) of 1.395 for \targ\ is near the maximum value of 1.4 for a reliable astrometric solution \cite{Gaia21aa}.

%%%%%%%%%%%%%%%%%%%%%%%%%%%%%%%%%%%%%%%%%%%%%%%%%%%%%%%%%%%%%%%%%%%%%%%%%%%%%%%%%%%%%%%%%%%%%%%%%%%%%%%%%%%%%%%%%%%%%%%%%%%%%%%%%%%%%%%%%%%%%%%%%%%%%

\section*{Summary and conclusions}

\targ\ is a dEB containing two F-type stars on a 1.90~d circular orbit. The system shows shallow (0.12 and 0.11 mag) partial eclipses which were discovered using the \hip\ satellite. We used TESS light curves from two sectors and published RVs from T08 to determine its physical properties. The partial eclipses make a solution of the light curve alone poorly determined, but the addition of a spectroscopic light ratio was sufficient to reach a determinate solution. The resulting radius measurements are relatively imprecise (2.3\%) due to this, and in comparison with the mass measurements (0.5\%). Our measured distance to the system is in reasonable agreement with that from \gaia\ DR3.

We compared the masses, radii and \Teff s of the stars to predictions from the {\sc parsec} stellar evolutionary models \cite{Bressan+12mn}. The models provide a match to these properties to within the 1$\sigma$ errorbars for an age of 800--900~Myr and a slightly supersolar fractional metal abundance of $Z=0.020$ (where the solar value is $Z=0.017$).

We also found the eclipses to arrive 11~min later than expected in the TESS light curves. Checks turned up no evidence for this being due to instrumental or data reduction issues, so it may be an astrophysical effect. The system should be monitored for eclipse timing variations caused by a possible third body. We also found residual systematics in the light curve which we attribute to weak starspots rotating synchronously with the orbit. Twenty-four observations with the \gaia\ Radial Velocity Spectrograph \cite{Cropper+18aa} yielded a double-lined spectroscopic orbit for the system which is in partial agreement with the ground-based results from T08. Future observations with \gaia\ should allow the addition of more RV measurements to this analysis, plus direct access to the \gaia\ spectra for checking the discrepancy found for one of the two stars.

%%%%%%%%%%%%%%%%%%%%%%%%%%%%%%%%%%%%%%%%%%%%%%%%%%%%%%%%%%%%%%%%%%%%%%%%%%%%%%%%%%%%%%%%%%%%%%%%%%%%%%%%%%%%%%%%%%%%%%%%%%%%%%%%%%%%%%%%%%%%%%%%%%%%%

\section*{Acknowledgements}

% We are grateful to Steve Overall and an anonymous referee for useful comments on a draft of this work.
We thank the anonymous referee for a quick and positive report.
This paper includes data collected by the TESS\ mission and obtained from the MAST data archive at the Space Telescope Science Institute (STScI). Funding for the TESS\ mission is provided by the NASA's Science Mission Directorate. STScI is operated by the Association of Universities for Research in Astronomy, Inc., under NASA contract NAS 5–26555.
This work has made use of data from the European Space Agency (ESA) mission {\it Gaia}\footnote{\texttt{https://www.cosmos.esa.int/gaia}}, processed by the {\it Gaia} Data Processing and Analysis Consortium (DPAC\footnote{\texttt{https://www.cosmos.esa.int/web/gaia/dpac/consortium}}). Funding for the DPAC has been provided by national institutions, in particular the institutions participating in the {\it Gaia} Multilateral Agreement.
The following resources were used in the course of this work: the NASA Astrophysics Data System; the SIMBAD database operated at CDS, Strasbourg, France; and the ar$\chi$iv scientific paper preprint service operated by Cornell University.

%%%%%%%%%%%%%%%%%%%%%%%%%%%%%%%%%%%%%%%%%%%%%%%%%%%%%%%%%%%%%%%%%%%%%%%%%%%%%%%%%%%%%%%%%%%%%%%%%%%%%%%%%%%%%%%%%%%%%%%%%%%%%%%%%%%%%%%%%%%%%%%%%%%%%

% \bibliographystyle{obsmaga}
% \bibliography{jkt}

%%%%%%%%%%%%%%%%%%%%%%%%%%%%%%%%%%%%%%%%%%%%%%%%%%%%%%%%%%%%%%%%%%%%%%%%%%%%%%%%%%%%%%%%%%%%%%%%%%%%%%%%%%%%%%%%%%%%%%%%%%%%%%%%%%%%%%%%%%%%%%%%%%%%%
\end{document}